\documentstyle[aps,preprint,psfig]{revtex}
\tighten
\input epsf

\begin{document}
\draft
\title{Resonant Andreev Tunneling in Strongly Interacting Quantum Dots}
\author{Rosario Fazio$^{(1,2)}$ and Roberto Raimondi$^{(3)}$}
\address{ 
 $^{(1)}$Istituto di Fisica, Universit\`a di Catania \& INFM, 
	 viale A. Doria 6, 95129 Catania - Italy$^{\ast}$\\
 $^{(2)}$Centre de Reserches sur les tr\`es Basses Temperatures - CNRS,
	 B.P. 166X, 38042 Grenoble, Cedex 9, France\\
 $^{(3)}$Dip. di Fisica "E. Amaldi", Universit\`a di Roma3 \& INFM,
	Via della Vasca Navale 84, 00146 Roma, Italy\\
}
\date{\today}
\maketitle
\begin{abstract}
We study resonant Andreev tunneling through a strongly interacting 
quantum dot connected to a normal and to a superconducting lead.
We obtain a formula for the Andreev current and apply it to 
discuss  the linear and non-linear transport in the nonperturbative
regime, where the effects of the Kondo resonance on  
the two particle tunneling arise. In particular we notice an 
enhancement of the Kondo anomaly in the $I-V$ characteristics due to 
the superconducting electrode. 
\end{abstract} 
 
The way a superconducting electrode affects the  electrical resistance
of a small mesoscopic region via Andreev tunneling, has been the subject
of an impressive research activity over the last years~\cite{Hekking94}. 
This has lead to the discovery of a number of new phenomena. Examples are 
the zero-bias anomalies in Normal-Insulating-Superconducting (NIS)
structures~\cite{Kastalsky91} or temperature (and voltage)
re-entrant behaviour in the conductance of a small metallic wire in contact 
with a superconducting lead~\cite{Charlat96}. Several of these phenomena 
have been successfully  explained by models of non-interacting 
electrons~\cite{Zaitsev90,Volkov93,Hekking93,Nazarov94,Beenakker94,Claughton96}. 
Electron-electron interaction strongly modifies the two-particle tunneling. 
In small capacitance junctions, for example, it leads to the Coulomb blockade of Andreev
tunneling~\cite{Hekking&Hergenrother}, whereas in a  Luttinger liquid-Superconducting
structure anomalous $I-V$ characteristics appear due to the excitation of the 
low lying modes~\cite{Fisher&Takane}.

Electrical transport via resonant tunneling in strongly interacting systems 
has been intensively investigated because of the  possibility  to study the 
behavior of an Anderson-like impurity model.
The prototype model is a Quantum Dot (QD), which mimics the impurity level,
coupled by tunnel junctions to two non-interacting leads. 
When the dot level is far below the Fermi energy of the leads, 
the formation of a spin singlet between the impurity spin and the conduction
electrons gives rise to a many-body resonance at the Fermi energy.
Such a resonance, known as Abrikosov-Suhl or Kondo resonance,
manifests in a peak in the density of states at the Fermi level and leads  
to a perfect transmission at zero temperature~\cite{Glazman&Ng}.
The two reservoirs, differently from the case of the Anderson impurity in 
metals, can be kept at different chemical potentials. Nonlinear transport
through a QD provides then a beatiful tool to study a Anderson impurity out 
of equilibrium~\cite{meir93}. The nonlinear $I-V$ characteristics and  the 
Kondo resonance in nonequilibrium conditions have been studied by a variety
of methods: the equation of motion for the Green's function~\cite{meir93},
variational procedure~\cite{Ng93}, a newly developed diagrammatic 
technique~\cite{Schoeller94}, and the $1/N$ expansion~\cite{Hettler95}. 
Experimental evidences of Kondo-like correlations have been reported in 
metal point contact~\cite{Ralph94} and very recently in SET 
transistors~\cite{Goldhaber97}.
Zero bias anomalies have also been reported recently in experiments 
involving superconductor/Anderson-insulator interfaces, though their
physical interpretation is still under debate~\cite{frydman97}. 

Remarkably a QD offers the, yet unexplored, possibility to study a quantum
impurity coupled to  two different types of electronic systems, 
such as superconductors and normal metals at the same time.
The development of superconductor-semiconductor
integration technology should allow to study the signatures of the Kondo 
effect in the Andreev tunneling. 

In this Letter we will study  Resonant Andreev Tunneling (RAT) in a
QD coupled to a normal and to a superconducting reservoirs as shown 
schematically in Fig.~\ref{fig1}.
Differently from the conventional problem of a Kondo impurity in a 
superconductor~\cite{Borkowski94} here the low lying excitations 
are still present due to the presence of the normal electrode.
  
The model Hamiltonian for the N-QD-S system under consideration can be 
written as
\begin{equation}
H=H_N +H_S+H_D+H_{T,N}+H_{T,S}
\label{model}
\end{equation}
where 
$
H_N=\sum_{k,\sigma} \epsilon_k c^{\dagger}_{N,k,\sigma} c_{N,k,\sigma}
$
, 
$
H_S=\sum_{k,\sigma} \epsilon_k c^{\dagger}_{S,k,\sigma} c_{S,k,\sigma}+
\sum_k (\Delta c^{\dagger}_{S,k,\uparrow} c^{\dagger}_{S,-k,\downarrow}
+ c.c.)
$
and
$
H_D=\epsilon_{d} d^{\dagger}_{\sigma}d_{\sigma}+Un_{d,\uparrow}
n_{d,\downarrow}
$
are the Hamiltonians of the normal lead, the superconducting lead 
($ \Delta $ is the superconducting gap) and the dot respectively.
The single particle energy $\epsilon_d$ is double degenerate in the 
spin index $\sigma$ and the interaction is included through the on-site 
repulsion $U$.
The position of the dot level can be modulated by an external gate voltage.
The tunneling between the leads and the dot is described by 
$
H_{T,\eta}=
\sum_{k,\sigma} (V_{\eta ,k,\sigma} c^{\dagger}_{\eta,k,\sigma,} d_{\sigma}
+ c.c. )
$
where $\eta =N,S$ and $V_{\eta ,k,\sigma}$ is the tunneling amplitude.

The average current, which for convenience we compute in the normal 
electrode, is given by
$
I =
2e{\sl Im}\sum_k <\Psi_{N,k}^{\dagger}
{\hat {\tau}}_z{\hat {H}}_k\Phi>   
$,
where we have adopted the Nambu notation 
(the hat  indicates matrices in the Nambu space): 
$\Psi_{N,k} =( c_{N,k,\uparrow},
c^{\dagger}_{N,-k,\downarrow})$ and
$\Phi = (d_{\uparrow} , d_{\downarrow}^{\dagger})$. ${\hat H}_k$ is a 
diagonal matrix
with elements $H_{11}=V_{N, k,\uparrow}$,$H_{22}=V^*_{N, -k,\downarrow}$. 

By means of the  Keldysh technique, as employed in Ref.~\onlinecite{Meir92},
the current $I$ can be rewritten in the form
\begin{equation}
\label{current3}
I = ie\int_{-\infty}^{\infty}
{{d\epsilon }\over {2\pi}} \Gamma_N {\bf Tr}
\lbrace {\hat {\tau}}_z  {\hat {G}}^R(\epsilon)    
[ 
 {\hat {\Sigma}}^R(\epsilon) {\hat {f}}_N (\epsilon) -
 {\hat {f}}_N (\epsilon){\hat {\Sigma}}^A(\epsilon)+
{\hat {\Sigma}}^<(\epsilon)]{\hat {G}}^A(\epsilon ) \rbrace 
\end{equation}
where ${\hat {G}}^{R(A)}$, ${\hat {G}}^<$ 
are the retarded (advanced) and
the lesser Green's functions of the dot
(for example ${\hat {G}}^{R}(t)=
-i\theta (t)<\lbrace \Phi (t),\Phi^{\dagger} (0)\rbrace >$).
In deriving eq.(\ref{current3}),
the relation ${\hat {G}}^<={\hat {G}}^R{\hat {\Sigma}}^<{\hat  {G}}^A$ 
has been used. 
Here $\Gamma_N (\epsilon ) =2 \pi \sum_k |V_{N,k,\sigma }|^2
\delta (\epsilon -\epsilon_k )$. For the sake of simplicity,
from now on we will neglect the 
spin and energy dependence of the tunneling matrix elements.
The diagonal matrix ${\hat {f}}_N$ has elements 
$f_{N,11}=f((\epsilon +eV )/T)$ and 
$f_{N,22}=1-f((-\epsilon +eV )/T)$ 
if the normal electrode is kept  at a chemical potential 
which differs by $eV$ from that of the superconductor 
($f(x)$ is the Fermi function and $V$ the voltage drop) .

To obtain a suitable expression for the current, we need to evaluate 
the lesser self-energy.    
In the non-interacting case,
${\hat {\Sigma}}^R(\epsilon)$ can be computed exactly and it 
is expressed in terms of the retarded and advanced self-energies as
$
{\hat {\Sigma}}^<_0 (\epsilon)=-\sum_{\eta =N,S} 
({\hat {\Sigma}}_{0,\eta}^R(\epsilon) {\hat {f}}_{\eta} (\epsilon) -
 {\hat {f}}_{\eta} (\epsilon){\hat {\Sigma}}_{0,\eta}^A(\epsilon)).
$
In the interacting case, we generalize Ng's ansatz to the present 
case~\cite{Ng96}. The lesser and greater self-energies are assumed 
to be of the form 
\begin{equation}
\label{intself}
{\hat {\Sigma}}^< ={\hat {\Sigma}}^<_0 {\hat {A}} 
\;\;\;\;\;\; , \;\;\;\;\;\;
{\hat {\Sigma}}^> ={\hat {\Sigma}}^>_0 {\hat {A}}
\end{equation}
where ${\hat {A}}$ is a matrix to be determined by the condition
${\hat {\Sigma}}^<-{\hat {\Sigma}}^>={\hat {\Sigma}}^R-{\hat {\Sigma}}^A$. 
This ansatz  is exact both in the non-interacting limit, $U=0$, and in the 
absence of  superconductivity, $\Delta =0$.
Moreover it guarantees automatically a current conserving 
scheme for any given approximation procedure to evaluate the retarded Green's
function. 
As a result we obtain
$
{\hat {\Sigma}}^<={\hat {\Sigma}}^<_0 ({\hat {\Sigma}}^R_0 -{\hat {\Sigma}}^A_0)^{-1}
({\hat {\Sigma}}^R-{\hat {\Sigma}}^A) \;\;\; .
$
The  expression for the current can be greatly simplified
in the relevant limit  $U,\Delta \gg k_BT,V$.
In this case  the  non-interacting self-energy due to
the superconducting lead ${\hat {\Sigma}}^{R(A)}_{0,S}$ is real and purely
off-diagonal whereas   that due to the normal lead,
${\hat {\Sigma}}^{R(A)}_{0,N}$, is diagonal. 
As a consequence, 
we obtain the  following form for the Andreev current through a QD 
\begin{equation}
\label{current4}
I =  ie\int_{-\infty}^{\infty}
{{d\epsilon }\over {2\pi}} \Gamma_N \nonumber \\
 {\bf Tr} \lbrace {\hat {\tau}}_z  {\hat G}^R(\epsilon)\left[
 {\hat {\Sigma}}^R(\epsilon), {\hat {f}}_N (\epsilon) \right]
 {\hat G}^A(\epsilon ) \rbrace \;\;\; .  
\end{equation}
This is the first central result of this work. It generalizes to the case
of a strongly interacting dot  the formula  valid
in the non interacting case~\cite{Beenakker92}  and
 allows to study the  transport through a 
N-QD-S under non-equilibrium situations and in the nonperturbative regime.

The presence of a superconducting lead introduces, besides $U$, another
energy scale in the problem, the energy gap $\Delta$.  
One is often interested to voltages and temperatures well below 
the energy scale set by $\Delta$ and   could be  tempted  to take the 
$\Delta\rightarrow\infty$ from the outset.
In an interacting quantum dot, the $\Delta\rightarrow\infty$ limit 
cannot be taken, in contrast to the non-interacting case.  
In order to have coherent Andreev scattering,
the two electrons enter the superconductor  without double occupying
the QD (which is forbidden in the $U\rightarrow\infty$
limit). This can happen only on a time scale of the order $1/\Delta$. 
We will then consider the case $U \gg \Delta \gg T,V$~\cite{limits}.

The generalization of the decoupling scheme for ${\hat G}^R$ in the 
presence of superconductivity is conveniently done by
rewriting the superconducting lead Hamiltonian, $H_S$, 
in terms of quasiparticles operators by means of a Bogoliubov 
transformation.
The reason for transforming to the quasiparticles basis is dictated
by the type of approximations introduced in the equation of motion
approach. In fact, one may generalize to the case of a superconducting
lead, the decoupling usually introduced in the normal case. Such 
decoupling, which neglects correlations in the lead, is done in terms
of the equilibrium number of quasiparticles. 
The final expression for the matrix
Green's function of the dot is~\cite{eom,details}  
\begin{equation}
\label{green}
\left( 
\begin{array}{c c}
\epsilon -\epsilon_d -\sigma_{N,11} (\epsilon )-\sigma_{S,11}(\epsilon)
&\Gamma_S /2 \\
 \Gamma_S /2 &  \epsilon +\epsilon_d -\sigma_{N,22}(\epsilon )-
\sigma_{S,22}(\epsilon)\\
\end{array}
\right)
{\hat G}  
=  
\left( 
\begin{array}{c c}
1-<n_{\downarrow}> & 0\\
0 & 1-<n_{\uparrow}>\\
\end{array}
\right)
\end{equation}
where $\sigma_{N,11}$ is the  $U=\infty$-limit self-energy considered
in Ref.\cite{meir93}
\begin{equation}
\label{nself}
{\sigma}_{N,11}(\epsilon) = -i{{\Gamma_N}\over {2}}+
\sum_k |V|^2{{f_N(\epsilon_k )}\over {\epsilon -\epsilon_k +i0^+}}
\end{equation}
and ${\sigma}_{S,11}$ is the corresponding self-energy due to the
superconducting electrode
\begin{equation}
\label{sself}
{\sigma}_{S,11}(\epsilon) =- 
\sum_k |V|^2{{v_k^2}\over {\epsilon +E_k +i0^+}}
\end{equation}
($v_k^2=1/2(1 - \epsilon_k/E_k )$ and
$E_k=\sqrt{\epsilon_k^2+|\Delta|^2}$). The self-energy for the hole
propagator can be obtained using the property 
${\sigma}_{N(S),22}(\epsilon) = {\sigma}_{N(S),11}^{*}(-\epsilon)$.
The self-energy  ${\sigma}_{S,11}$ 
is weakly energy dependent due to the low energy 
cutoff provided by $\Delta$ (in what follows, we consider
$\omega_c\gg\Delta \gg T,\epsilon$  with $\omega_c$ the bandwidth).
At energies much smaller than the gap quasiparticles present in the dot
cannot decay by tunneling into the superconductor 
(as the imaginary part of the diagonal 
elements of the self-energy ${\hat \sigma}_S$ vanishes).
As a result the contribution of the self energy due to the superconducting 
lead ${\sigma}_{S,11}$ simply  shifts the dot level to the new value
$\tilde{\epsilon }_d \approx \epsilon_d +(\Gamma _S/2\pi )
\ln \omega_c/\Delta$.
The divergence of the level energy renormalization with $\Delta$ reveals
that the process occurs via a virtual state in which a quasiparticle 
is created in the superconductor. It is however due to the weak
(logarithmic) nature of this divergence that RAT can can be observed 
in any realistic situation.  
Note also that the off-diagonal element, $\Gamma_{S}/2$, 
entering eq.(\ref{green}), has the same form of the non interacting case
apart from an overall phase-shift of  $\pi$.
This $\pi$ phase-shift comes from a relative cancellation between the 
non-interacting and  interacting off-diagonal self-energies. 
Substituting the espression of the QD's Green's function 
in Eq.(\ref{current3}) we get the desidered result
\begin{equation}
\label{current5}
I(V) =\int^{\infty}_{-\infty}~d\epsilon 
{{f(\epsilon -eV)-f(\epsilon +eV)}\over {2e}} G_{NS}(\epsilon )
\end{equation}
with 
\begin{equation}
\label{conductance}
G_{NS}(\epsilon )={{4e^2}\over h} 
{{2(\Gamma_N \Gamma_S )^2 ((\Gamma_{1,N} +\Gamma_{2,N})/2\Gamma_{N}))}\over
{\left[4 (\epsilon-\epsilon_1)(\epsilon+\epsilon_2)-\Gamma_{1,N}\Gamma_{2,N}
-\Gamma_{S}^2\right]^2 +4\left[\Gamma_{1,N}(\epsilon+\epsilon_2)+
   \Gamma_{2,N}(\epsilon-\epsilon_1)\right]^2} }
\end{equation}
\noindent
where 
$\epsilon_{1(2)} =\epsilon_d + {\sl Re}{\hat {\sigma}}_{N,11(22)}
+{\sl Re}{\hat {\sigma}}_{S,11(22)}$,  
$\Gamma_{1(2),N}=-2{\sl Im}{\hat {\sigma}}_{N,11(22)}$. 
Eqs.(\ref{current4},\ref{conductance}) are the main results of the present 
paper.

The spectral function $G_{NS}(\epsilon )$ associated with the resonant Andreev 
tunneling is plotted in Fig.\ref{fig2} for various bias voltages 
(for comparison, the non-interacting case is shown in the inset).
Several features are worth noticing.  First, two peaks at 
$\pm \tilde{\epsilon}$ are due to particle and hole bare levels.
Note that in the interacting case the bare level energy includes
the renormalization due to the superconducting electrode self-energy
as discussed above. Second, at low temperatures a Kondo peak develops
at the Fermi energy.  Quite remarkably, 
at finite positive (negative) voltages the Kondo peak shifts pinned to the 
Fermi level of the normal metal while a small kink develops at negative 
(positive) voltages. At  finite voltages  hole and particle energies differ by 
$2eV$, and while the electron (hole) is on resonance for positive (negative)
voltage, the Andreev reflected hole (electron) is off resonance with respect 
to the shifted Fermi level. 
Compared to the N-QD-N case, here there is no peak splitting. 
Third, the Kondo peak remains rather pronounced even in the 
nonequilibrium situation. The differences between the N-QD-N
and the  N-QD-S cases, can be traced back to the fact that the
superconducting electrode acts simply as a boundary condition, even in the 
nonequilibrium situation, the Kondo resonance being achieved through the 
tunneling into the normal electrode. 
For this  reason  the Kondo peak is shifted but  not suppressed.

The differential conductance, for various temperatures,
is shown in Fig.~\ref{fig3}. In the low temperature regime the zero bias 
anomaly associated with the Kondo effect is clearly seen. We note 
that the Kondo peak survives up to temperatures of the order of $\Gamma /4$, 
about one order of magnitude larger than the N-QD-N case. 
This effects is shown in the inset of  Fig.~\ref{fig3} where the zero bias 
anomaly  for the N-QD-S (present work) is compared with the corresponding 
quantity in the N-QD-N, Ref.~\cite{meir93}. The anomaly in the RAT 
clearly survives in a temperature regime where is already absent for normal
tunneling.
This may be useful for the experimental detection of the effect. 

In this Letter we studied the RAT in a Normal metal - Quantum Dot - 
Superconductor device. An explicit form of the current through the device is 
obtained. The analysis of the $I-V$ characteristics has been carried out 
in the limit $U \gg \Delta \gg T,V$, where two electrons tunnel almost 
indipendently since the onsite Coulomb
repulsion (which is the largest energy scale in the problem) prohibits the double
occupancy of the Dot. In this case  the Kondo effect enhances the 
Andreev conductance at low temperatures. 
In the opposite limit, $ \Delta \gg U \gg T,V$, different processes dominate
the transport and a suppression of the Andreev tunneling is expected. 
The results were obtained by generalizing, in the presence of superconductivity,    
an ansatz due to Ng and by means of the equation of motion method to determine the 
Green's function of the Dot. While the expression for the current in
Eq.~(\ref{current4}) holds under general circumstances, 
the results for the spectral function and the differential conductance
are quantitatively valid for temperatures larger than the Kondo temperatures
and only qualitatively valid for lower temperatures~\cite{meir93,Ng96}.
We then expect that, even though a more refined 
treatment is necessary for more quantitative results in the Kondo region,
all qualitative features of our findings will survive.
The combined effect of a finite  $ \Delta $ and $ U $ is currently 
being investigated and it will be a subject of a forthcoming publication.

\acknowledgments
We thank F.W.J. Hekking for many fruitful discussions.
We acknowledge the financial support of INFM under the PRA-project
"Quantum Transport in Mesoscopic Devices" and EU TMR programme
(Contract no. FMRX-CT 960042).

\begin{figure}
\caption{The system under consideration. A Quantum Dot coupled by tunnel 
barriers to a Normal and to a Superconducting electrodes. The position of the
level in the dots can be tuned by means of the gate voltage $V_g$}
\label{fig1}
\end{figure}
\begin{figure}
\caption{The spectral density for the two particle 
tunneling $G_{NS}(\epsilon)$  is plotted for various bias voltages ( 
$V=0$ solid line, $ V=\pm 0.005$ dotted line, 
$ V=\pm 0.01$ dashed line, $ V=\pm 0.015$ dot-dashed line,  
$\tilde{\epsilon}_d = -0.07$, $\Gamma_S=\Gamma_N = 0.02$ and $T = 0.0001$, in units
of the bandwidth $\omega_c$). 
In the inset the non interacting case is shown for comparison.}
\label{fig2}
\end{figure}
\begin{figure}
\caption{The differential conductance of the NQDS device, in units of
$4e^2/h$, is plotted for 
different temperatures ($T=0.0001$ solid line, $ T = 0.001$ dotted line, 
$ T= 0.01$ dot-dashed line, $\tilde{\epsilon_d} = -0.04$, 
$\Gamma_S=\Gamma_N = 0.02$, in units
of the bandwidth $\omega_c$). In the inset the zero bias anomaly,  detected
by measuring the relative change of the differential conductance $G(V)$ between
zero and high voltage ($\Delta G /G \equiv (G(0)-G(-0.04))/G(0)$), is shown for the 
N-QD-S (triangles ) and for the  N-QD-N (filled circles) \protect\cite{meir93}}
\label{fig3}
\end{figure}
\newpage
{\epsfxsize=12cm\epsfysize=10cm\epsfbox{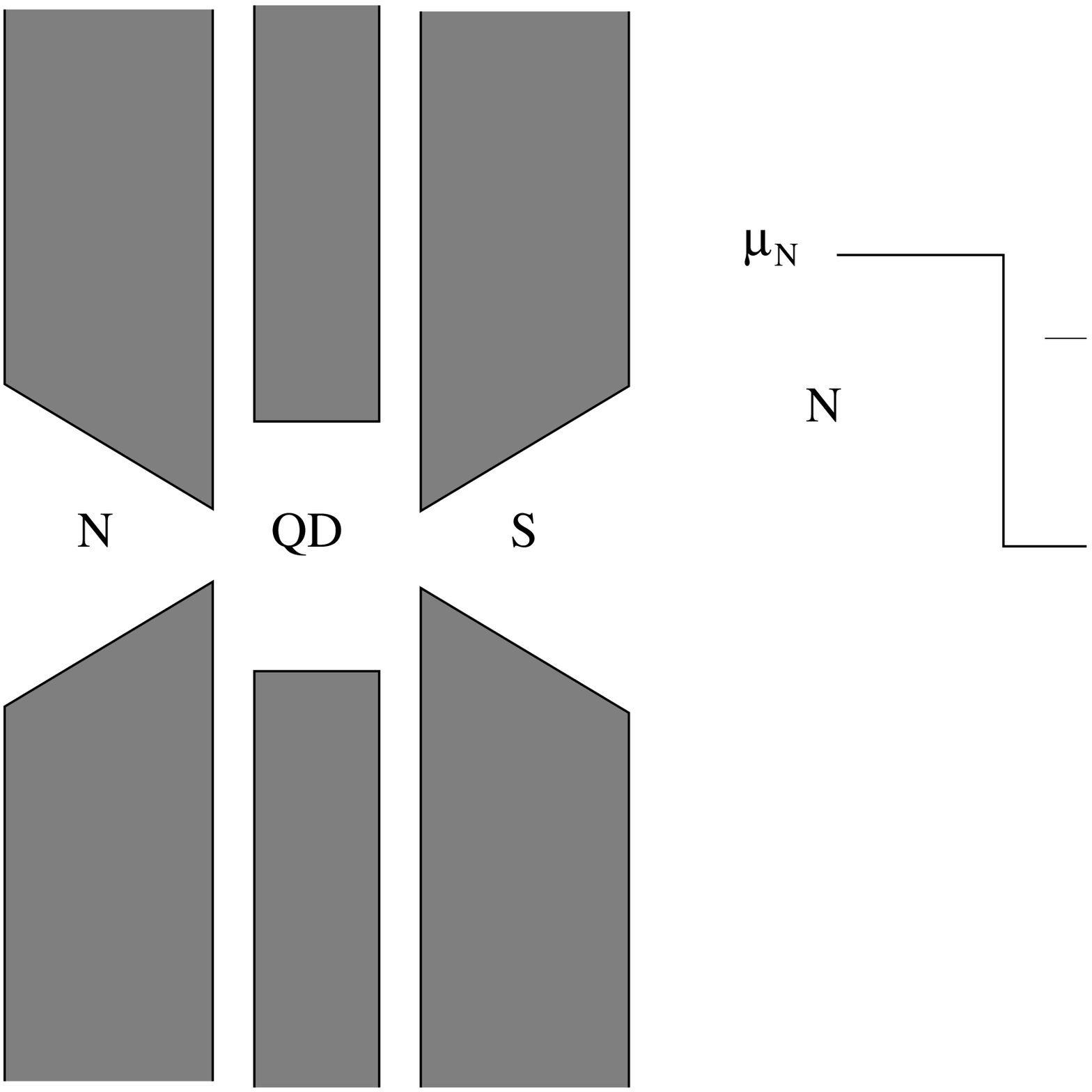}}
\newpage
{\epsfxsize=12cm\epsfysize=10cm\epsfbox{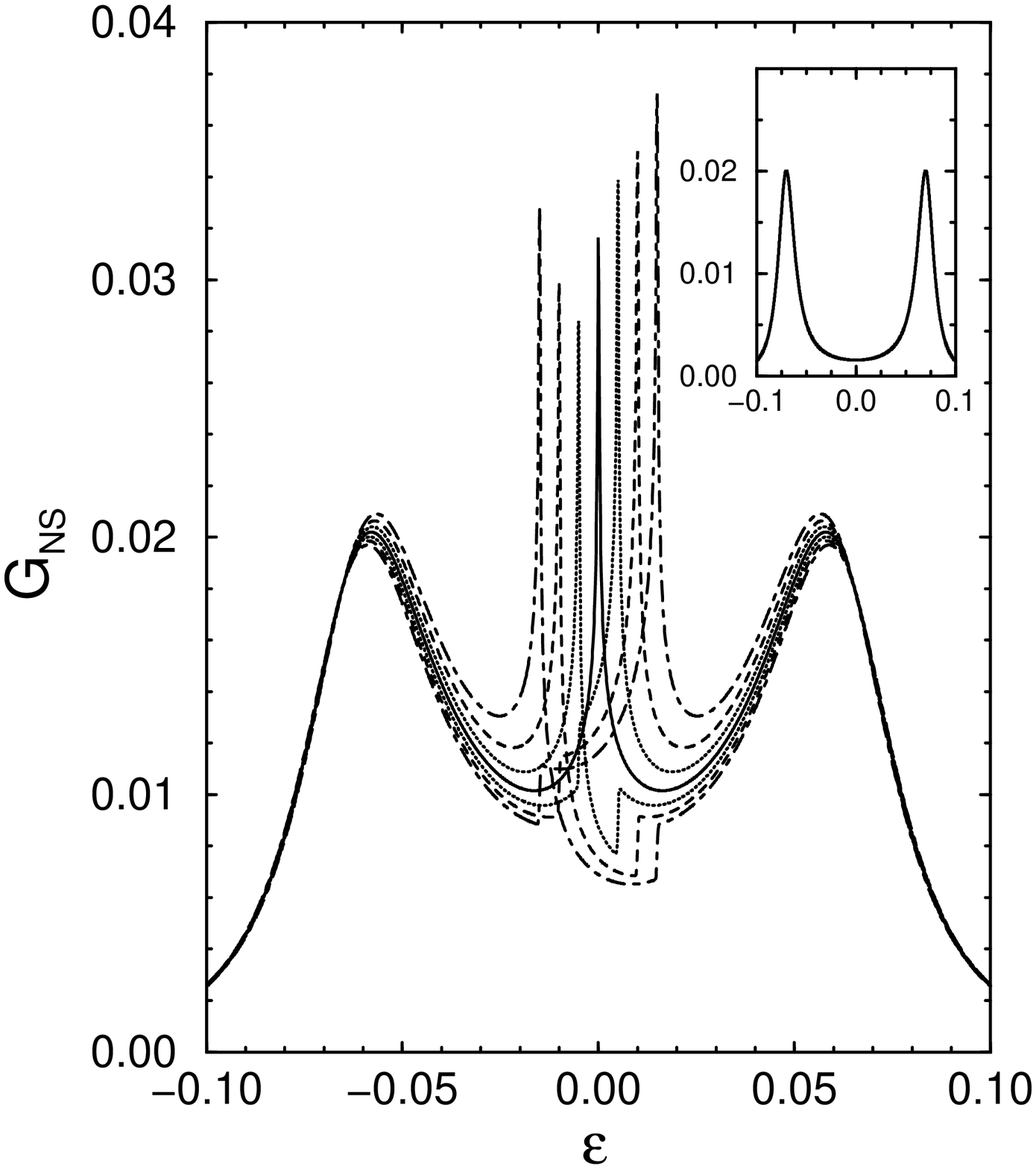}}
\newpage
{\epsfxsize=12cm\epsfysize=10cm\epsfbox{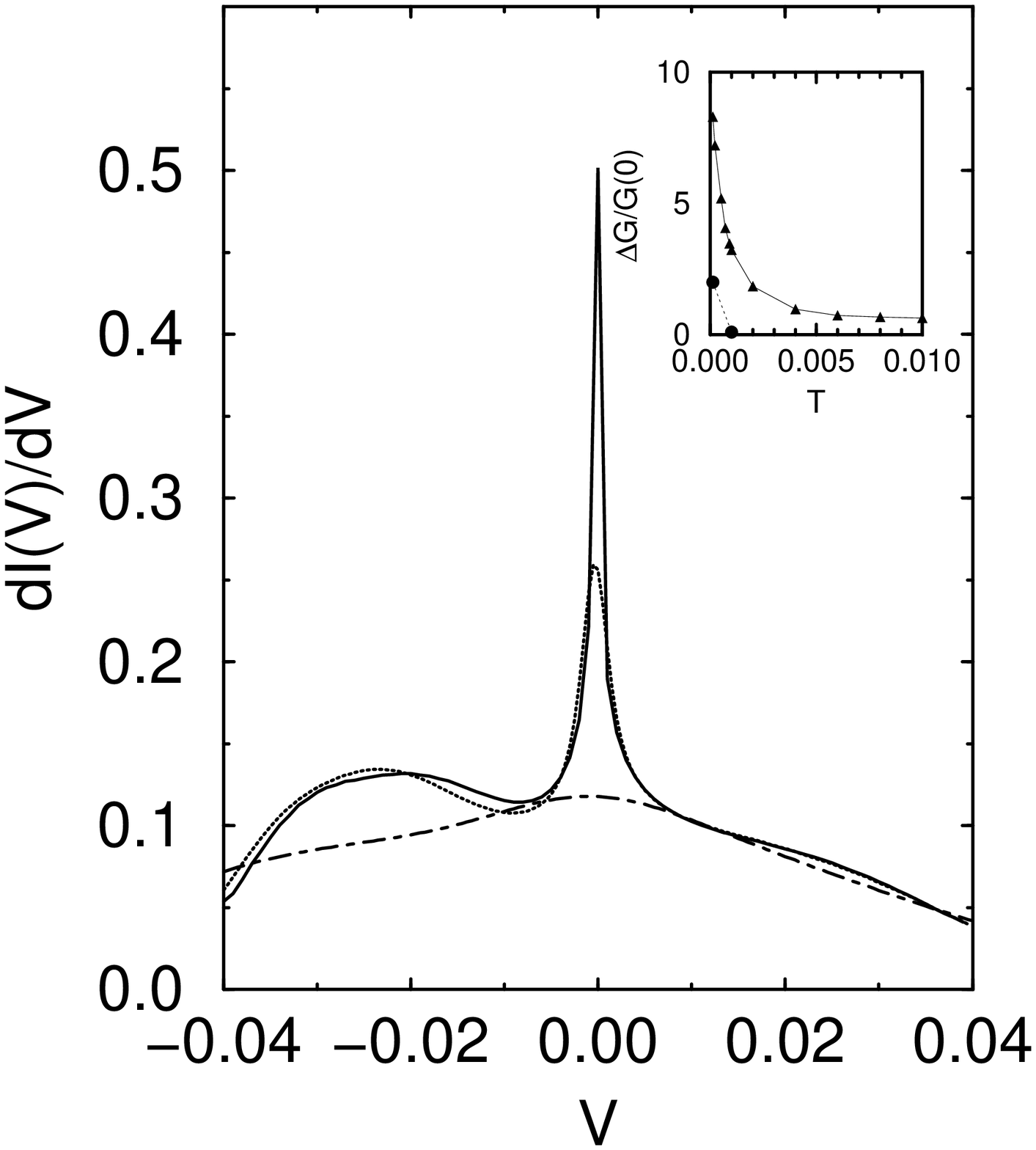}}


\begin{references}
\bibitem[\mbox{$\ast$}]{Permanent address.} Permanent address.
\bibitem{Hekking94}
	{\it Mesoscopic Superconductivity}, Proceedings of the NATO ARW,
	F.\,W.\,J. Hekking, G. Sch\"on, and D.\,V. Averin, eds.,
	Physica B {\bf 203} (1994);
	C.W.J. Beenakker in {\em Mesoscopic
	Quantum Physics}, edited by E. Akkermans,  G. Montambaux,
	and J.-L Pichard (North Holland, Amsterdam) 1995.
\bibitem{Kastalsky91}
	A. Kastalsky, A. W. Kleinsasser, L. H. Greene, R. Bhat,
	F. P. Milliken, and J. P. Harbison, Phys. Rev. Lett.
	{\bf 67}, 3026 (1991).
\bibitem{Charlat96}
	P. Charlat, H. Courtois, Ph. Gandit, D. Mailly,
	A.F. Volkov, and B. Pannetier, Phys. Rev. Lett.
	{\bf 77}, 4959 (1996).
\bibitem{Zaitsev90} A. V. Zaitsev Sov. Phys. JETP {\bf 51}, 41 (1990).
\bibitem{Volkov93} A. F. Volkov, A. V. Zaitsev, and T. M. Klapwijk
	 Physica C {\bf 210}, 21 (1993).
\bibitem{Hekking93}F.W.J. Hekking and Y.V. Nazarov Phys. Rev. Lett. 
	{\bf 71}, 1625 (1993).
\bibitem{Nazarov94} Y. V. Nazarov Phys. Rev. Lett. {\bf 73}, 1420 (1994).
\bibitem{Beenakker94} C. W. J. Beenakker, B. Rejaei, and J. A. Melsen
	 Phys. Rev. Lett. {\bf 72}, 2470 (1994).
\bibitem{Claughton96} N. R. Claughton, R. Raimondi, and 
	C. J. Lambert Phys. Rev. B {\bf 53}, 9310 (1996).
\bibitem{Hekking&Hergenrother} F.\,W.\,J. Hekking, L.\,I. Glazman,
	K.\,A. Matveev, and R.\,I.
	Shekhter, Phys. Rev. Lett. {\bf 70}, 4138 (1993);
	J.M. Hergenrother, M.T. Tuominen and M. Tinkham, {\em ibid}
	{\bf 72}, 1742 (1994).
\bibitem{Fisher&Takane} M.P.A. Fisher, Phys. Rev. B {\bf 49},
	14550 (1994); Y. Takane and Y. Koyama, J. Phys. Soc. Japan, 
	{\bf 66}, 119 (1997).
\bibitem{Glazman&Ng} L. I. Glazman and M. E. Raikh JETP Lett.
        {\bf 47}, 452 (1988);
	T. K. Ng and P. A. Lee Phys. Rev. Lett.
        {\bf 61}, 1768 (1988). 
\bibitem{meir93} Y. Meir, N. S. Wingreen, and P. A. Lee 
	Phys. Rev. Lett. {\bf 70}, 2601 (1993).
\bibitem{Ng93} Tai Kai Ng  Phys. Rev. Lett. {\bf 70}, 3635 (1993).
\bibitem{Schoeller94}
	H. Schoeller and G. Sch\"on, Phys. Rev. B {\bf 50}, 18436 (1994).
\bibitem{Hettler95}M.~H. Hettler and H. Schoeller, Phys. Rev. Lett. 
	{\bf 74},4907 (1995).
\bibitem{Ralph94} D. C. Ralph and R. A. Buhrman Phys. Rev. Lett.
        {\bf 72}, 3401 (1994).
\bibitem{Goldhaber97} D. Goldhaber, H. Shtrikman, D. Mahalu, D.
	Abush-Magder, U. Meirav and M.A. Kastner, unpublished 
	(report cond-mat/9707311). 
\bibitem{frydman97} A. Frydman and Z. Ovadyahu Phys. Rev. B {\bf 55},
         9047 (1997).
\bibitem{Borkowski94} L.S. Borkowski and P.J. Hirschfeld, J. Low Temp. Phys.
	{\bf 96}, 185 (1994), and references therein.
\bibitem{Meir92}Y. Meir and N.~S. Wingreen, Phys. Rev. Lett. {\bf 68},
	2512 (1992). 
\bibitem{Ng96} Tai Kai Ng  Phys. Rev. Lett. {\bf 76}, 487 (1996).
\bibitem{Beenakker92}
	C.W.J. Beenakker, Phys. Rev. B {\bf 46}, 12481 (1992);
	N. R. Claughton, M. Leadbeater, and C. J. Lambert
	J. Phys.: Condensed Matter {\bf 7}, 8757 (1995). 
\bibitem{limits} In order to appreciate this point, let us consider the  
	Hamiltonian of an isolated quantum dot, plus a term which models 
	the Andreev  scattering at the boundary 
	$H_{T,S}=(t_{A } d^{\dagger}_{\uparrow}d^{\dagger}_{\downarrow}+ c.c.)$.
	This problem can be solved exactly and the off-diagonal element of the 
        inverse Green's function reads  
	$-t_A \left(1+{U(\epsilon +\epsilon_d + U)}/ 
	({\epsilon^2 -(\epsilon_d +U)^2 -t_A^2})\right)$
	which vanishes in the $U\rightarrow\infty$ limit. Then, in the limiting
	sequence $\Delta\rightarrow\infty$ first, $U\rightarrow\infty$ after,
	the induced superconductivity in the dot is completely suppressed.
	The ansatz discussed in the paper, Eq. \protect\ref{intself}, is exact
	in this soluble case.
\bibitem{eom} As discussed in Ref.\protect\onlinecite{meir93}, the 
	equation of motion method does not provide quantitative results in
	the Kondo region, however it gives the qualitative feature both
	in and out of equilibrium case.
\bibitem{details}Details of the derivation will be published elsewhere,
	R. Fazio and R. Raimondi (unpublished).

\end{references}
\end{document}